\documentclass[aps,pra,twocolumn,floatfix,10pt,final]{revtex4}
\pdfoutput=1
\makeatletter
\def\@dotsep{4.5}
\makeatother

\usepackage{amsmath}
\usepackage{graphicx}
\usepackage{amssymb}
\usepackage{epstopdf}
\begin{document}
\title{Charge-Transfer Photodissociation of Adsorbed Molecules via Electron Image States}
\author{E.T. Jensen}
\email{ejensen@unbc.ca}
\affiliation{\\Physics Department, University of Northern British Columbia\\3333 University Way, Prince George B.C. V2N 4Z9, Canada}

\date{\today}      

\begin{abstract}
The 248nm and 193nm photodissociation of submonolayer quantities of CH$_3$Br and CH$_3$I adsorbed on thin layers of {\it n}-hexane indicate that the dissociation is caused by dissociative electron attachment from sub-vacuum level photoelectrons created in the copper substrate. The characteristics of this photodissociation-- translation energy distributions and coverage dependences show that the dissociation is mediated by an image potential state which temporarily traps the photoelectrons near the {\it n}-hexane--vacuum interface, and then the charge transfers from this image state to the affinity level of a co-adsorbed halomethane which then dissociates. 
\end{abstract}
\maketitle


\section{Introduction}
One of the recurring themes in surface photochemistry on metal surfaces is the almost ubiquitous contribution that photoelectrons make to the observed photochemical processes. In particular, many phenomena have been interpreted as being the consequence of {\it hot photoelectrons\/}- those photoelectrons having energies between the Fermi and the vacuum levels of the system. In nanosecond timescale photochemistry on noble metal substrates, the system has sufficient time for many cycles of electron excitation, scattering and decay, so that the resultant electron energy distribution that results has a form $n(E)\propto (E-E_{Fermi})^{-2.5}$ at energies well above $E_{Fermi}$\cite{Weik:1993}. This rather smooth energy distribution can interact with the relevant electron affinity levels of the adsorbate system. If one is particularly interested in the dissociation of molecules via dissociative electron attachment (DEA), then one would expect that the convolution of the $n(E)$ with the $\sigma_{DEA}\left(E\right)$ will yield a flux of desorbing species that arises from a relatively wide range of photoelectron energies (roughly the width of $\sigma_{DEA}(E)$ for energies a few eV above $E_{Fermi}$). 

One can inquire if there are mechanisms that will modify this nascent energy distribution-- in particular, can the photoelectron energy distributions be narrowed to be more selective in the dissociation of adsorbed species. A related question has to do with electron transport in thin molecular films and how the electrons in these films can interact with other species- again, are there physical systems that will be highly selective in the interactions between co-adsorbed molecules? One approach was first detailed on theoretical grounds by Rous\cite{Rous:1995} who considered how an incoming electron might couple with the electron affinity levels of a $N_2$ molecule near a metal surface to vibrationally excite the molecule. It was shown by application of a KKR calculation that the electron affinity level $\sigma_{vib}(E)$ would display a strong enhancement by resonance with electron image states of the surface system.

In order to form such an image state and have an image state lifetime sufficient to yield a significant enhancement, one requirement is that the surface have a bandgap at the relevant electron energy (typically E$\sim$-0.5eV relative to $E_{vacuum}$ for the {\it n\/}=1 state), which does indeed occur for a number of low index metal surfaces. A practical complication is that these image states on clean metal surfaces have been found to be very sensitive to contamination effects-- rather small concentrations of some adsorbates on metal surfaces have been found to open decay channels that suppress the image state \cite{Donath:1992}. A different approach is to take advantage of image states that can be formed at dielectric surfaces- these were first described in detail by Cole and Cohen\cite{Cole:1969} and were subsequently observed using microwave spectroscopy on {\it l\/}-He surfaces\cite{Grimes:1974}. More recent work on image states at dielectric thin films has been in the context of characterizing the energies and lifetimes of these states using femtosecond two-photon photoemission (fs-2PPE) spectroscopy. In general terms, molecules having negative electron affinities (i.e. repulsive for low energy electrons) in thin layers form a bandgap that encompasses the image state energies (i.e. the energy region just below $E_{vacuum}$). In qualitative terms, the decay of these electron image states is dominated by the transmission probability through the barrier, a barrier that increases the lifetimes roughly exponentially with film thickness. There is now a substantial body of work on the effect of thin films interacting with intrinsic metal image states for both the case of available conduction band states at the image state energies (image state resonance, e.g. Kr and Xe on Cu(100)\cite{Berthold:2004}) and pure image electron barrier layers (e.g. thin alkane layers\cite{Harris:1997}) on various metal surfaces. There has also been work on coupling of a negative ion resonance for O$_2$ with an image state resonance formed for Xe/Cu(111), in which the O$_2$ anion state couples to the Xe conduction band, with a strong layer dependence for the anion state lifetime\cite{Hotzel:1998}.

\begin{figure}
\resizebox{1.00\columnwidth}{!}{
\includegraphics{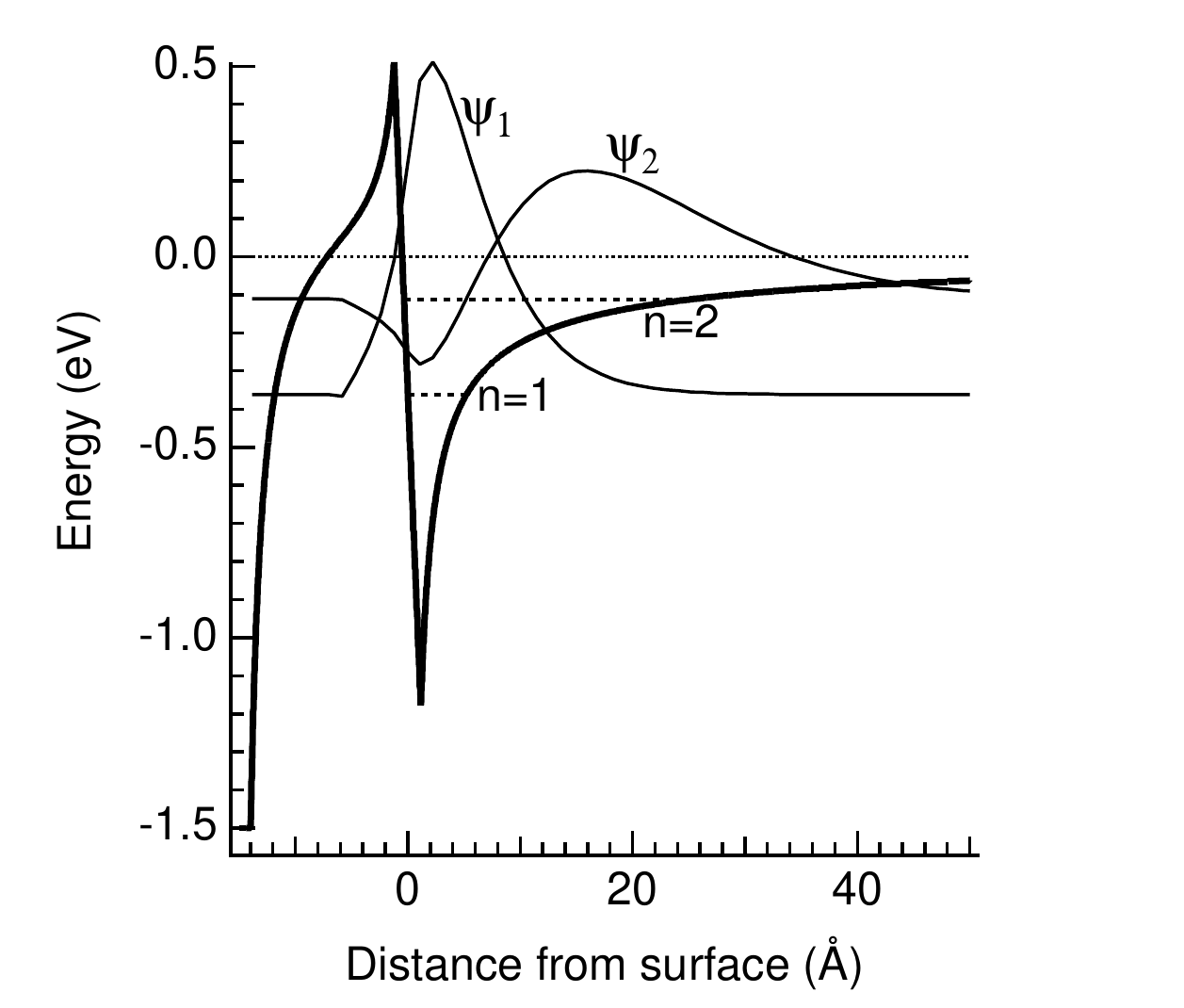}}
\caption{Schematic image potential and 2 lowest bound states for approximately 4ML (15\AA) of {\it n\/}-hexane on Cu(110) obtained using a dielectric continuum model. The {\it n\/}-hexane--vacuum interface is at {0\AA}  and the Cu(110) surface is located at -15\AA. The negative electron affinity of {\it n\/}-hexane (EA$\approx$-0.2eV) creates a barrier at the surface which can trap electrons in the image states. The Cu(110) substrate has no bandgap at the image state energies, so electrons that tunnel through the {\it n\/}-hexane barrier can enter unoccupied metal states directly.}
\label{Image_States}
\end{figure}

For the present case, a schematic image potential at a 4ML {\it n\/}-hexane on Cu(110) is shown in Fig. {\ref{Image_States}}. We have used the modified dielectric continuum model of Hotzel {\it et al\/}\cite{Hotzel:1999} to describe the image potential at the {\it n\/}-hexane--vacuum interface, and physical values similar to those chosen in recent fs-2PPE studies of this system (for {\it n\/}-hexane, EA=-0.20eV, $\epsilon_r=2.0$)\cite{Harris:1997}. The lowest lying image states have binding energies of 0.36eV ({\it n\/}=1) and 0.11eV ({\it n\/}=2). In the case of the Cu(110) substrate used in this work there is no intrinsic surface bandgap near $\overline{\Gamma}$ at these energies, so that image state electrons that tunnel through the {\it n\/}-hexane barrier layer will simply enter Cu conduction band states. In the experiments described below, a submonolayer of `detector' molecules (CH$_3$Br or CH$_3$I) are adsorbed on top of the {\it n\/}-hexane layer. If the dissociative attachment cross section for these adsorbed molecules has a substantial overlap with the image state energy and the image state has sufficient lifetime, we expect that there would be image state mediated DEA observed. Fig. {\ref{Image_States}} also shows that the largest spatial overlap between the image states and an adsorbed molecule will occur for the {\it n\/}=1 image state-- for higher {\it n\/} image states, the wavefunction becomes increasingly weighted away from the surface. From fs-2PPE experiments, it is also found that the {\it n\/}=1 image state has a substantially higher population than the {\it n\/}=2 and higher states\cite{Lingle:1996}, so one would expect dissociation to occur primarily from the {\it n\/}=1 state.

For CH$_3$Br and CH$_3$I in the gas phase, DEA occurs for low energy electrons with a threshold electron energy related to the curve crossing of the neutral ground state potential with the anion state\cite{Schramm:1999, Wilde:2000}. The photodissociation of these halomethanes at metal surfaces and the role of photoelectron mediated DEA has been studied by a variety of workers\cite{Zhou:1991, Ukraintsev:1992, Dixon-Warren:1993}. When these molecules are adsorbed at a metal or dielectric surface, the anion state is shifted downward in energy and the threshold electron energy for DEA decreases and the survival probability for the anion increases, so that in general the peak DEA cross section is at lower electron energy and has a substantially larger magnitude. These effects are well known and have been previously reported in the literature (e.g. see Ref. \cite{Sanche:1995}). In the case of CH$_3$Br and CH$_3$I, the lowering of the anion potential by $\sim1eV$ will result in the threshold electron energy for DEA at or close to 0eV and the peak DEA cross section at very low electron energy. At the surface, these correspond to electron energies below the vacuum level\cite{Ayotte:1997,Jensen:2007}, and can be accessed by the hot photoelectrons discussed above and, as will be discussed in the present work, will also have energetic overlap with electron image states at the surface.

\section{Experimental Methods}
Experiments were performed using an ultra-high vacuum surface science and photochemistry apparatus that has been described previously\cite{Jensen:2005}. We have utilized both Cu(100) and Cu(110) single crystal substrates in this work. The samples are heated to 915K by electron bombardment for cleaning and cooled to $\sim$92K using liquid nitrogen, the temperature at which the experiments were performed. Sample cleanliness and order is monitored by Auger electron spectroscopy and low energy electron diffraction measurements. Neutral products from surface photodissociation travel 185mm to pass through a 4mm diameter aperture to a differentially pumped Extrel quadrupole mass spectrometer with an axial electron bombardment ionizer. The sample to ionizer distance is 203mm. Ions created in the ionizer then travel to the quadrupoles and are mass selected, in the present experiments using m/q=15amu (CH$_3^+$). Selected ions are detected by a conversion dynode and electron multiplier in pulse counting mode and the pulses are then passed through a fast preamplifier. The particle arrival pulses are time recorded using a multichannel scaler (MCS) that begins counting prior to the initiating laser pulse, with a typical time-of-flight (TOF) spectrum using 1000 1.0$\mu$s time bins, and summing the counts from 1000 laser pulses.

The laser pulses ($\sim$5ns duration) are produced by a small excimer laser (MPB PSX-100) operating at 20Hz. Both unpolarized and linearly polarized light has been used in this work. To create polarized light, the beam passes through a birefringent MgF$_2$ crystal to separate p- and s-polarized light, which can then be directed at the sample. Unless otherwise stated, p-polarized laser pulses were used in the present study. Various laser wavelengths are produced by selection of different laser gases: 193nm (ArF), 248nm (KrF) and 308nm (XeCl). Pulse energies varied for the different laser wavelengths, but the fluence on the sample were less than 1mJ/cm$^2$. The laser pulses were collimated using a 4mm diameter aperture and were unfocused on the sample.  The laser light is incident upon the sample at an angle of 45$^\circ$ from the mass spectrometer axis-- when the Cu sample is aligned to collect mass fragments along the surface normal, the light is incident at 45$^\circ$. Angular distributions of the photofragments were measured by rotating the sample with respect to the mass spectrometer axis, which also changes the incident angle of the laser light. In the present experiments the effect of changing the incident light angle is small and its consequence on photoelectron generation is well understood.

The liquids used for dosing in the present work: {\it n\/}-hexane (Aldrich $\geqslant$99.5\%) and CH$_3$I (Aldrich 99.5\%)- were transferred to a glass and teflon gas-handling system and degassed by freeze-pump-thaw cycles. The CH$_3$Br (Aldrich 99.5\%) was used as delivered from a lecture bottle. Room-temperature vapor pressures were used for gas dosing through a precision leak valve that backfilled the chamber. Gas doses  are described using uncorrected ion gauge readings. Gas doses that correspond to monolayer coverages were calibrated using temperature programmed desorption measurements or from prior experience with the selected molecule in other experiments.

\section{Results and Discussion}
\begin{figure}
\scalebox{0.50}{
\includegraphics{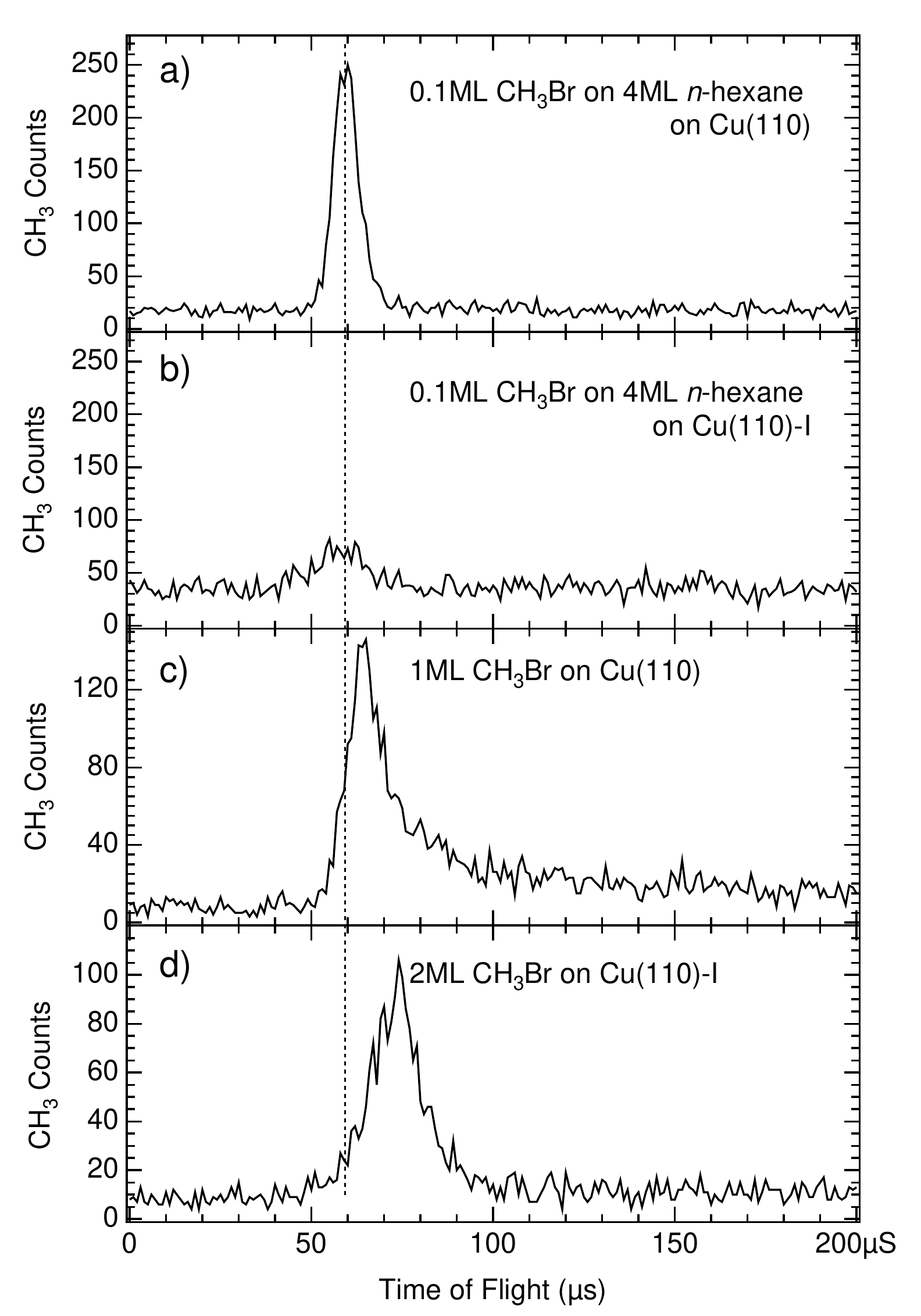}}
\caption{Time-of-flight spectra for CH$_3$Br obtained using 1000 pulses of 248nm light. In (a) the signal is from a submonolayer (0.1ML) of CH$_3$Br on top of 4ML {\it n\/}-hexane on the Cu(110) substrate. Spectrum (b) is obtained in a similar situation to (a), except that the substrate is c(2x2) Cu(110)-I. The spectrum (c) is obtained from 1ML of CH$_3$Br on Cu(110), and (d) is obtained from 2ML CH$_3$Br on Cu(110)-I. In the spectra (a)-(c), the CH$_3$ signal is collected along the surface normal, while in (d) the signal is collected 20$^\circ$ from the normal- in all these cases, these angles are the direction of maximum CH$_3$ yield.}
\label{CH3Br_TOF_Comparison}
\end{figure}

A set of TOF spectra using 248nm ($h\nu$=5.0eV) laser pulses for CH$_3$Br  adsorbed on Cu(110) are shown in Fig. {\ref{CH3Br_TOF_Comparison}}. In the gas-phase, CH$_3$Br has a low cross section for direct photodissociation at this wavelength, so it is expected that DEA by photoelectrons will be the dominant dissociation mechanism. This expectation is supported by the observed TOF spectra.  In Fig. {\ref{CH3Br_TOF_Comparison}}(a) the TOF spectrum for a submonolayer of CH$_3$Br adsorbed on a 4ML {\it n\/}-hexane layer on Cu(110) shows the single narrow feature that is ascribed to the CT-DEA via an image state intermediate. That this feature is caused by photoelectrons is supported by the spectrum of Fig. {\ref{CH3Br_TOF_Comparison}}(b), in which the substrate photoelectron yield is reduced due to the addition of atomic iodine to the Cu surface, which forms a c(2x2) Cu(110)-I surface\cite{Johnson:2000,Andryushechkin:2005}. This iodization largely suppresses the CH$_3$Br dissociation signal, though a small peak at the same flight time is just visible. That this small feature is observed at the same flight time (i.e. same CH$_3$ translational energy) supports the notion of an image state intermediate at the {\it n\/}-hexane surface, since one would expect that the photoelectron energy distribution would be altered by the iodization. In general, changes in the photoelectron energy distribution will be reflected in the TOF spectrum if the DEA occurs via a resonance at an energy where the photoelectron distribution is altered. In Figs.  {\ref{CH3Br_TOF_Comparison}}(c) and (d) the CH$_3$Br image state mediated dissociation to differs from that of 1ML CH$_3$Br on clean Cu(110) (Fig. {\ref{CH3Br_TOF_Comparison}}(c)) and 2ML CH$_3$Br on Cu(110)-I (Fig. {\ref{CH3Br_TOF_Comparison}}(d)). In these latter spectra, the peak of the TOF spectrum is shifted and the TOF peaks are broader than that of Fig. {\ref{CH3Br_TOF_Comparison}}(a). The absolute width in energy (FWHM) for the peak in Fig. \ref{CH3Br_TOF_Comparison}(a) is 0.18eV, while for Figs \ref{CH3Br_TOF_Comparison}(c) and (d) the widths are 0.28 and 0.24eV respectively. The relative peak widths ($\Delta E/E_{peak}$) are 0.32, 0.62 and 0.64 respectively, again highlighting the distinctly narrower translational energy distribution for dissociation via the image state electron mechanism. In particular, comparison of the spectra of Figs {\ref{CH3Br_TOF_Comparison}}(a) and (d) for CH$_3$Br on Cu(110)-I is telling, since this CH$_3$Br layer is known to be quite well orientationally ordered and can display very narrow TOF features\cite{Jensen:2006}. At 248nm, it appears that the width of the TOF feature for CH$_3$Br/Cu(110)-I in Fig.  {\ref{CH3Br_TOF_Comparison}}(d) is dominated by the range of photoelectron energies that cause DEA, resulting in CH$_3$ fragments of more widely varying kinetic energy. The narrow TOF spectrum of Fig. {\ref{CH3Br_TOF_Comparison}}(a) for CH$_3$Br/hexane/Cu(110) is the first piece of evidence that the DEA is dominated by a very narrow range of incident electron energies,  due to coupling with an image state that has captured a photoelectron which is then coupled to the CH$_3$Br affinity level for DEA.

\begin{figure}
\scalebox{0.50}{
\includegraphics{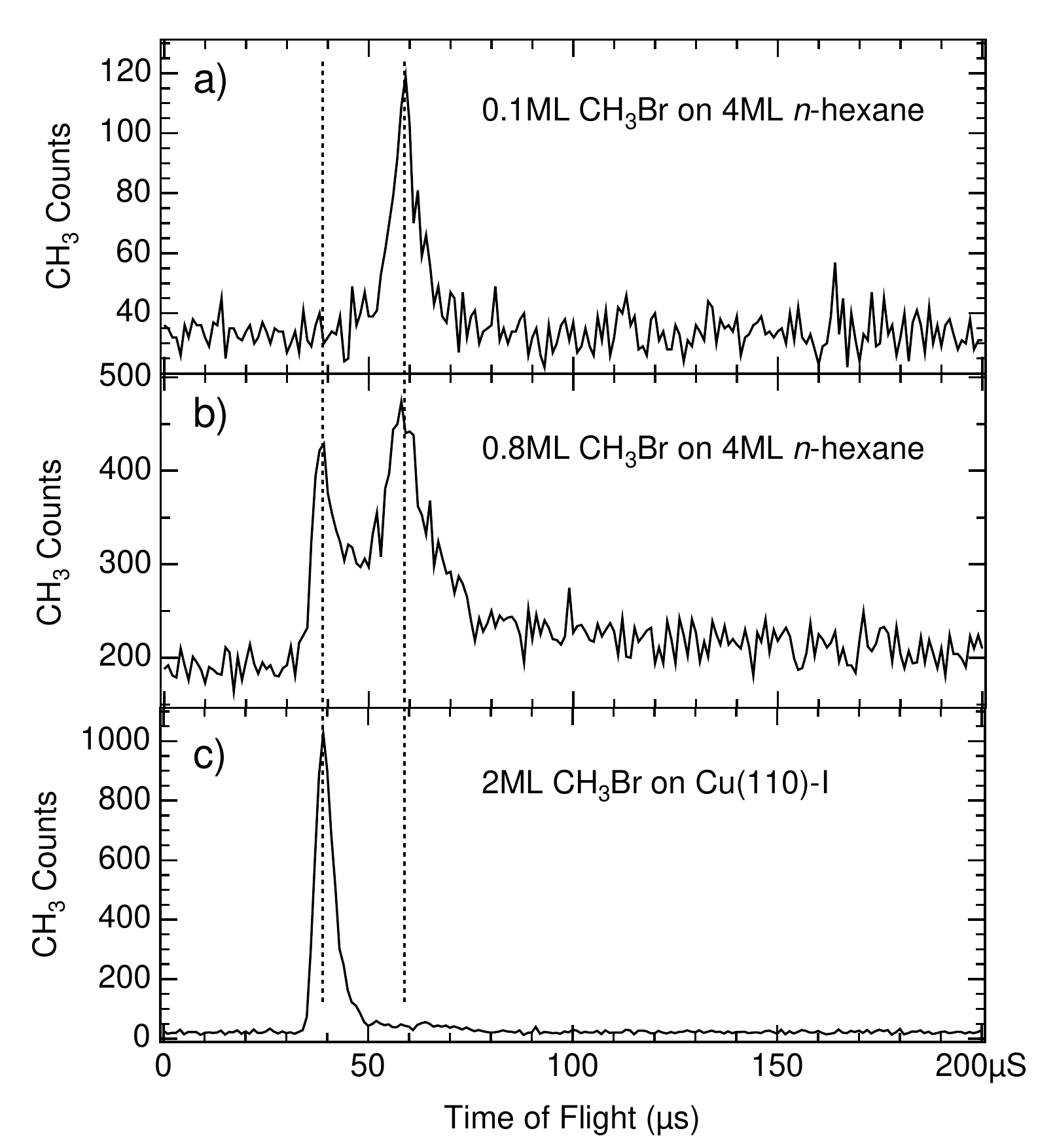}}
\caption{Time-of-flight spectra for adsorbed CH$_3$Br obtained using 193nm light. In (a) the signal is from 0.1ML CH$_3$Br on top of 4ML {\it n\/}-hexane on the Cu(110) substrate, and shows a single TOF feature due to image state mediated photodissociation. In spectrum (b), the CH$_3$Br coverage is increased to 0.8ML, and two TOF features are seen-- the same image state mediated dissociation as for (a) at 59$\mu s$ flight time and a feature due to direct photodissociation at 39$\mu s$. As a comparison, (c) shows a TOF spectrum for CH$_3$Br on Cu(110)-I in which the direct photodissociation process dominates and there is no image state mediated dissociation visible. }
\label{CH3Br_193nm_Comparison}
\end{figure}

Time-of-flight spectra for CH$_3$Br using 193nm ($h\nu$=6.40eV) are shown in Fig. {\ref{CH3Br_193nm_Comparison}}. In contrast to the 248nm case, at 193nm CH$_3$Br is known to have a large cross section for direct photodissociation in the gas-phase\cite{VanVeen:1985-1} and at surfaces\cite{Jensen:2006}. It is therefore somewhat surprising to find that for 0.1ML of CH$_3$Br on top of 4ML of {\it n\/}-hexane on Cu(110) that the same TOF distribution is seen using 193nm light as for 248nm (Fig. {\ref{CH3Br_193nm_Comparison}}(a)). The peak in this TOF spectrum is at exactly the same flight time as for the 248nm case, showing the same energy release in CH$_3$ translation in spite of the 1.40eV higher photon energy. That the position and width of this TOF feature is unchanged shows that the electrons responsible for this dissociation have the same energy distribution and is further support for the image state mediated mechanism of photodissociation. At 193nm, as the CH$_3$Br coverage in increased, the total CH$_3$ yield increases but a new feature at 44$\mu s$ flight time emerges (Fig. {\ref{CH3Br_193nm_Comparison}}(b)) and at higher CH$_3$Br coverage, dominates the spectrum. This feature is associated with the 193nm direct photolysis of CH$_3$Br, as it peaks at the same flight time as that seen in a previous study\cite{Jensen:2006}- a representative TOF spectrum is shown in Fig. {\ref{CH3Br_193nm_Comparison}}(c) for comparison. Thus even at a wavelength at which CH$_3$Br has a large direct photodissociation cross-section, the image state mediated dissociation dominates at low coverage. This is due to the large effective cross section for this process as the photoelectrons that are promoted to the image state are free-electron like parallel to the {\it n\/}-hexane surface, and can cause DEA in adsorbed species that are relatively distant from the location of the initial excitation. For example, for a trilayer of {\it n}-pentane on Ag(111), the measured $n$=1 image state lifetime is 17.6ps (and increasing by roughly $10\times$ per added monolayer)\cite{Harris:1997}, so that such an $m^*$=1 image state electron with $\sim$0.2eV of kinetic energy would have a mean free path length of roughly 4.6$\mu$m. At higher CH$_3$Br coverages (approaching 1 monolayer and above), the image state mediated process persists but the direct photodissociation mechanism becomes more competitive.

\begin{figure}
\scalebox{0.60}{
\includegraphics{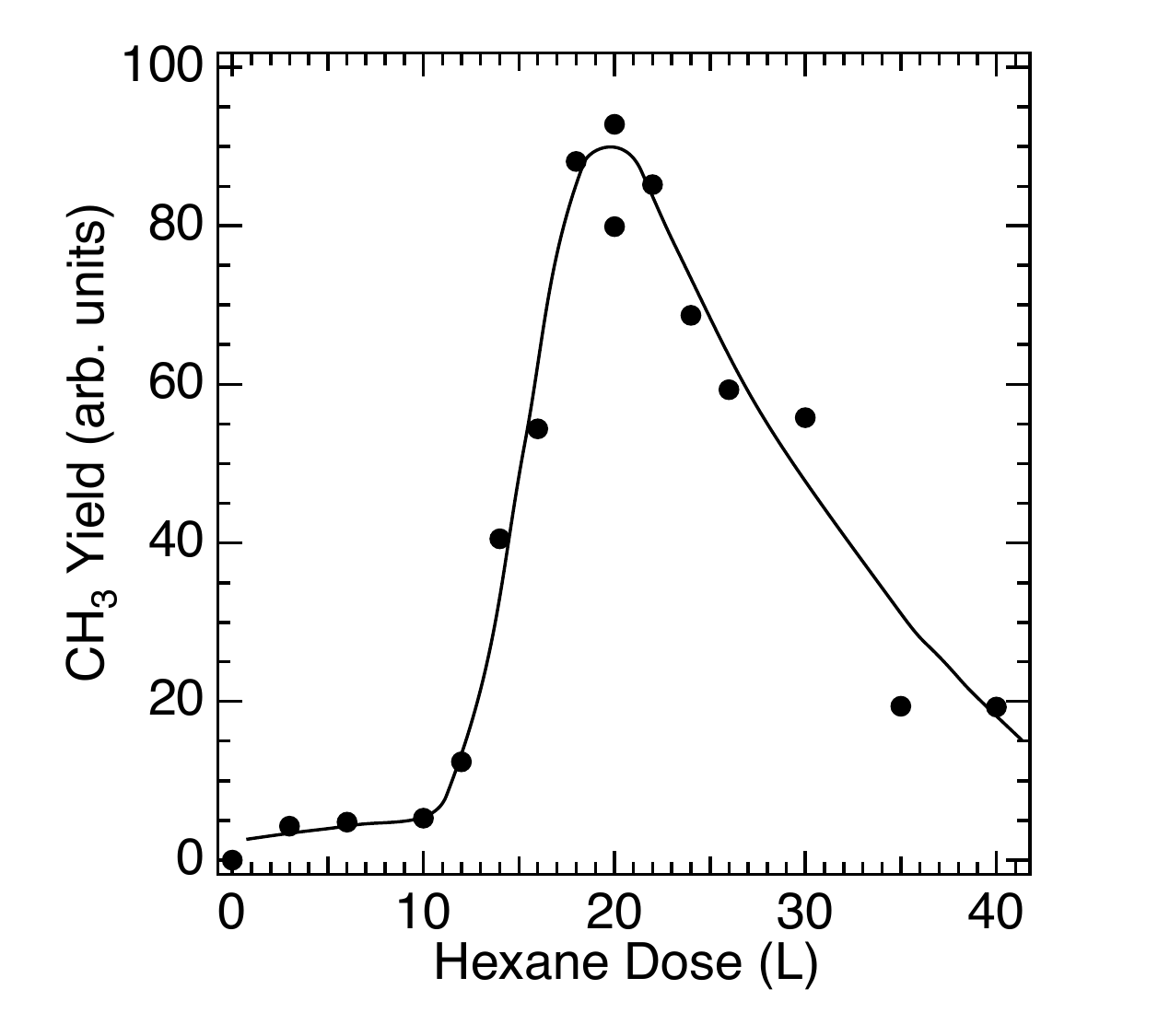}}
\caption{Measured yield of CH$_3$ from the 248nm photolysis of 0.1ML CH$_3$Br on top of a varying thickness of {\it n\/}-hexane on Cu(110). These data points are obtained from a series of individual TOF spectra, each the result of 1000 laser pulses, obtained in the surface normal direction. The solid curve is shown as a guide to the eye. }
\label{CH3Br_Vary_Hexane_Dose}
\end{figure}

When the thickness of the {\it n\/}-hexane ``barrier" layer is varied for a fixed submonolayer of CH$_3$Br, the CH$_3$ yield is found to vary as shown in Fig. {\ref{CH3Br_Vary_Hexane_Dose}}. From temperature programmed desorption (TPD) measurements made for {\it n\/}-hexane on Cu(110) and Cu(100), the completion of the first monolayer occurs at approximately 4.5L {\it n\/}-hexane dose. If it is assumed that the sticking coefficient is roughly constant for subsequent layers under our conditions (borne out by TPD measurements), we observe from Fig. {\ref{CH3Br_Vary_Hexane_Dose}} that very little CH$_3$ yield is obtained for 2ML or less of {\it n\/}-hexane. The CH$_3$ yield in Fig. {\ref{CH3Br_Vary_Hexane_Dose}} is observed to increase significantly as the {\it n\/}-hexane coverage increases beyond 2ML and reaches a maximum at approximately 4ML {\it n\/}-hexane thickness. At coverages beyond 4ML, the CH$_3$ yield is found to decrease rapidly. Similar measurements made using the Cu(100) substrate showed essentially identical behavior-- little or no CH$_3$ yield from 2ML or less of {\it n\/}-hexane and a rapid rise in yield to a maximum near 4ML {\it n\/}-hexane coverage. The decreasing CH$_3$ yield for film thicknesses beyond 4ML is not due to changes in the CH$_3$ angular distribution, as substantial angular changes were not observed. It is possible that the adsorbed submonolayer is increasingly subsumed in the {\it n\/}-hexane layers for higher coverage (e.g. at defects), though we do not see changes in the TOF spectrum peak shapes (i.e. translational energy distribution) or angular distributions. When Cu(100)-I or Cu(100)-Cl substrates are used, there are much lower yields (e.g. see Fig. \ref{CH3Br_TOF_Comparison}(b)) but the coverage for maximum CH$_3$ yield is observed to occur in the range of 1--2ML {\it n}-hexane.

In studies of image state lifetimes of similar molecular systems using fs-2PPE, it is found that the lifetime increases rapidly (roughly exponentially) with increasing layer thickness. The Cu(110) substrate used in the present work has no surface bandgap at the image state energies, so the decay of the populated image states will be dominated by the electron tunneling through the {\it n\/}-hexane barrier. In contrast, the Cu(100) surface has a prominent s-p bandgap around $\bar{\Gamma}$ for energies relevant to the image states. That we observe essentially identical photodissociation behavior on both surfaces indicates that the metal surface intrinsic bandgap is not playing a significant role in these experiments. The low yield of CH$_3$ for 2ML or less of  {\it n\/}-hexane suggests that the image state lifetimes are too short for the image state electrons to localize on the CH$_3$Br adsorption sites, and that for 3 to 4ML {\it n\/}-hexane coverage, the probability to locate and dissociate the CH$_3$Br becomes favorable. Another possibility is that the dissociating CH$_3$Br$^-$ temporary anion is quenched more rapidly for these thin {\it n\/}-hexane films, presumably by overlap with unoccupied metal states (i.e. autodetachment). Experience from previous DEA photodissociation experiments would suggest that this is likely more important on the single monolayer thickness films. The optimal $\sim$4ML {\it n\/}-hexane thickness for DEA from co-adsorbed CH$_3$Br seems to represent a compromise between the formation and population of the image state and the image state lifetime. For thicker {\it n\/}-hexane films ($>$4ML) the decreased CH$_3$ yield is most likely due to a decrease in the initial population of the {\it n\/}-hexane--vacuum interface image state, which becomes increasingly less probable as the requisite photoelectrons generated in the metal substrate cannot tunnel through the thick {\it n\/}-hexane layers. 

One more possibility to consider is that the CH$_3$Br dissociation is due to DEA from an interface electron state {\it inside} the {\it n\/}-hexane layer (similar to states identified in Ar thin films\cite{Rohleder:2005}), and the rapid rise in the 2ML--4ML region is due to the exclusion of this state from the thinner films. Sustaining such a state requires a surface bandgap for the metal substrate at the relevant energies and we observe no differences between the behavior on Cu(100) (which has such a gap at $\bar{\Gamma}$) and Cu(110) (which does not). Also when experiments were performed on Cu(100)-I or Cu(100)-Cl the maximum signal was seen for much thinner films (1--2ML) so we do not believe that such an interface electron state in {\it n\/}-hexane is responsible for the observed dissociation.

\begin{figure}
\scalebox{0.60}{
\includegraphics{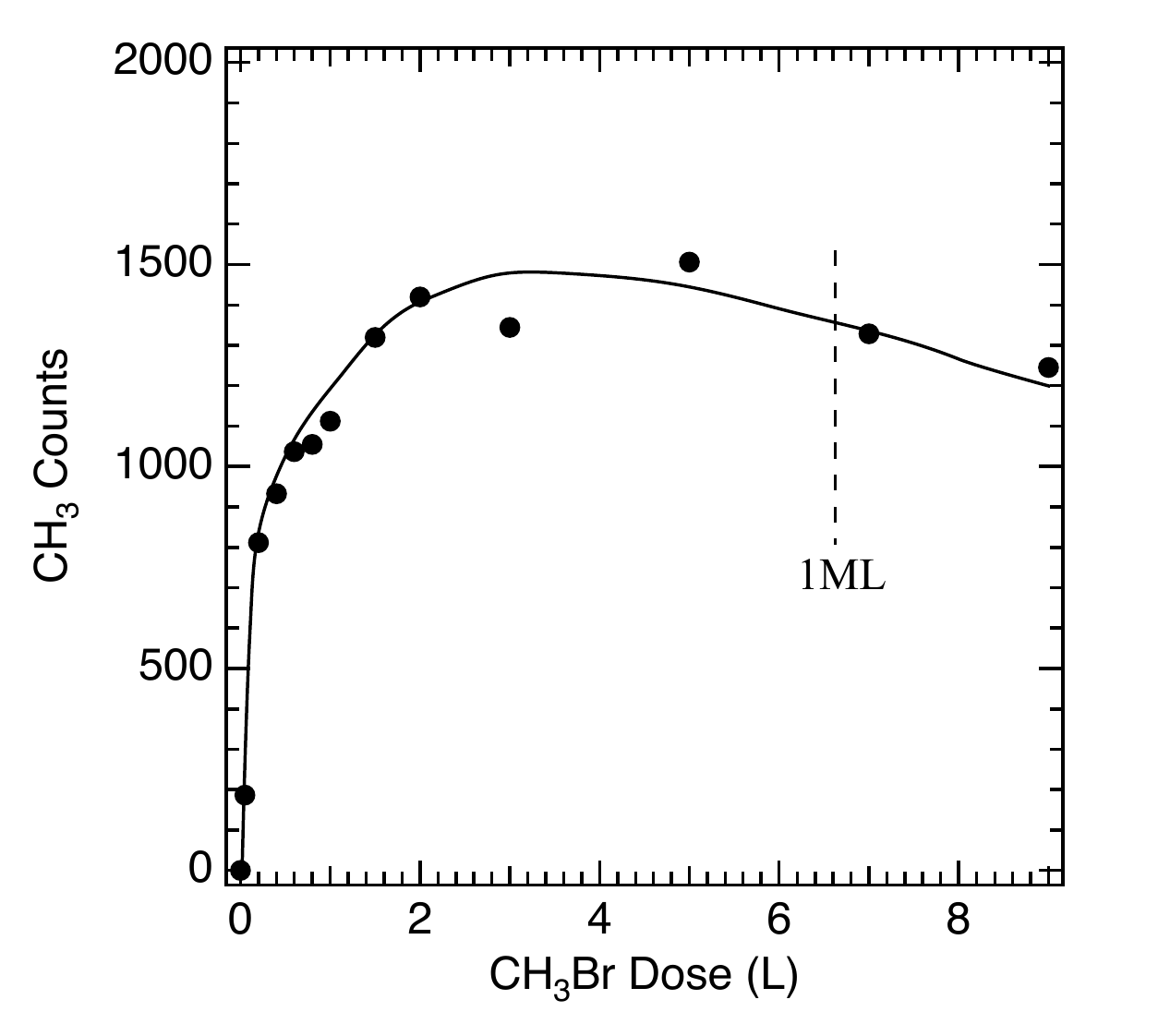}}
\caption{Yield of CH$_3$ from photolysis of varying amounts of CH$_3$Br adsorbed on $\sim$4ML {\it n\/}-hexane on Cu(110). The measured counts are detected in the surface normal direction from 1000 pulses of 248nm light. From TPD experiments on clean Cu(110), one nominal monolayer of CH$_3$Br corresponds to $\sim$6.5L CH$_3$Br dose. The large increase in CH$_3$ yield corresponds to the range 0--0.3ML CH$_3$Br coverage. }
\label{Vary_CH3Br_on_Hexane}
\end{figure}

If the observed photodissociation of adsorbed CH$_3$Br is mediated by an image state electron intermediate, then one expectation would be that this photodissociation would be very efficient for even a small submonolayer of the dissociating species. The image state electron at the thin film--vacuum interface, though constrained in the surface normal direction, would be expected to be mobile parallel to the surface as described above. The data of Fig. {\ref{Vary_CH3Br_on_Hexane}} support this notion- there is a very steep increase in the observed CH$_3$ yield for submonolayer CH$_3$Br on the 4ML {\it n}-hexane on Cu(110) system using 248nm light. From TPD and photochemistry experiments on clean Cu(110), we have found that a 6.5L dose of CH$_3$Br corresponds to one monolayer coverage, and for adsorption on {\it n}-hexane, one would anticipate that the sticking coefficient would be, at best, the same, so that the observed increase in the region of 0--2L dose corresponds to roughly 0--0.3ML coverage range. At higher coverages (above $\sim$0.5ML) the CH$_3$ angular distribution becomes broader, so the observed roughly constant yield at higher doses measured in the surface normal direction in Fig. {\ref{CH3Br_Vary_Hexane_Dose}} is not entirely reflective of the total yield, however the steep increase in yield at low coverage is not affected by this. From image state studies at clean metal surfaces with intrinsic surface bandgaps (e.g. using inverse photoemission), a sensitivity of the image state to surface contamination has been noted\cite{Donath:1992}. This surface contamination sensitivity does not seem to carry over to the image states created at thin {\it n}-hexane surfaces, the characteristic image state dissociation is observable even at monolayer coverages of the dissociating species (for example, see Fig. {\ref{CH3Br_193nm_Comparison}}(b)).

\begin{figure}
\scalebox{0.60}{
\includegraphics{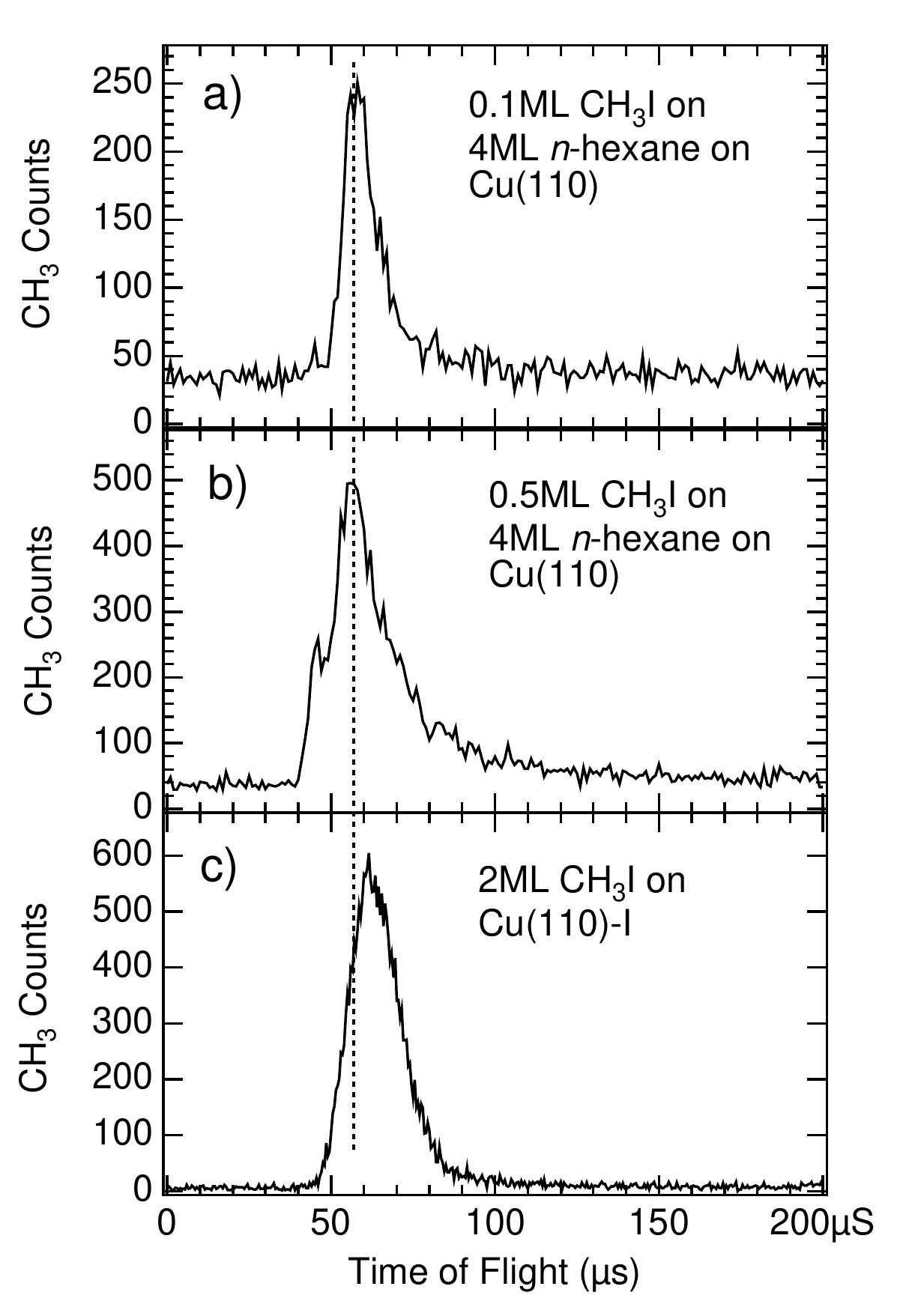}}
\caption{Time-of-flight spectra from CH$_3$I:  (a) from a submonolayer on 4ML {\it n\/}-hexane (248nm s-polarized light), (b) from 0.5ML CH$_3$I on 4ML {\it n\/}-hexane (248nm p-polarized light) and (c) from 2ML CH$_3$I on Cu(110)-I obtained using s-polarized 308nm light. }
\label{CH3I_TOF_Comparison}
\end{figure}

In addition to CH$_3$Br, we have also observed the same image state mediated dissociation occur for adsorbed CH$_3$I submonolayers. CH$_3$I has a larger cross section for DEA in the gas-phase than CH$_3$Br does, with a peak cross section of $>$100$\times 10^{-16}$cm$^2$ for $E<$60meV\cite{Schramm:1999}. In the adsorbed state, CH$_3$I has a large DEA cross section for very low energy electrons ($\sigma=6.8\times 10^{-16}$cm$^2$ at E$\sim$0eV when adsorbed on 10ML of Kr)\cite{Jensen:2007}, with the peak DEA cross section having apparently shifted to an energy below the vacuum level due to image interactions with the surface. A TOF spectrum obtained at 248nm for 0.1ML CH$_3$I on 4ML {\it n}-hexane on Cu(110) is shown in Fig. {\ref{CH3I_TOF_Comparison}}(a). In this case, the CH$_3$I TOF distribution is narrower than that seen for CT-DEA of adsorbed CH$_3$I in other cases. If the CH$_3$I coverage is increased, and p-polarized 248nm light is used, photodissociation by both CT-DEA and by direct photodissociation can be seen (Fig. {\ref{CH3I_TOF_Comparison}}(b)). Similar to the case for CH$_3$Br at 193nm, direct photodissociation can be seen in the small fast peak at 45$\mu s$ flight time.  Another contrast can be seen in Fig. {\ref{CH3I_TOF_Comparison}}(c) which shows a TOF spectrum obtained from 2ML CH$_3$I that is well ordered on a Cu(110)-I substrate at 308nm ($h\nu=4.02eV$) using s-polarized light. Under these conditions, the TOF spectrum is dominated by the DEA mechanism\cite{Jensen:2005} from subvacuum level photoelectrons. The TOF spectrum for the CH$_3$I on {\it n}-hexane is noticeably narrower than that obtained at the longer wavelength on Cu(110)-I. 

Although for CH$_3$I we observe that the CH$_3$ TOF distributions are not as narrow as those found for CH$_3$Br, they are still narrower than those observed in other situations. One possible explanation for the slightly broader CH$_3$ TOF distribution for CH$_3$I, as compared to CH$_3$Br is that CH$_3$I does have a substantially larger intrinsic (i.e. not image state mediated) DEA cross section, so that there may be an additional contribution to the observed TOF signal from a broader range of DEA photoelectron energies.  That DEA can dominate the CH$_3$I dissociation at 248nm is in itself surprising, since the cross section for direct photolysis at this wavelength is quite large. Similar to the case for CH$_3$Br at 193nm, for small submonolayer coverages of CH$_3$I at 248nm, the image mediated DEA dominates the photodissociation. Direct photolysis is found to be increasingly important at higher CH$_3$I coverage, especially when p-polarized light is present since the oriented CH$_3$I has a large cross section in this case\cite{Jensen:2005}. It has also been found that all of the same general findings made for CH$_3$Br are also found for CH$_3$I-- the pronounced yield variation for varying {\it n\/}-hexane spacer thickness, the large observed yields at low CH$_3$I coverage and very similar angular distributions of the CH$_3$ photofragments. The findings made for CH$_3$I complement those made for CH$_3$Br and lend further support to the proposed image state mediated dissociation mechanism. 

\begin{figure}
\scalebox{0.60}{
\includegraphics{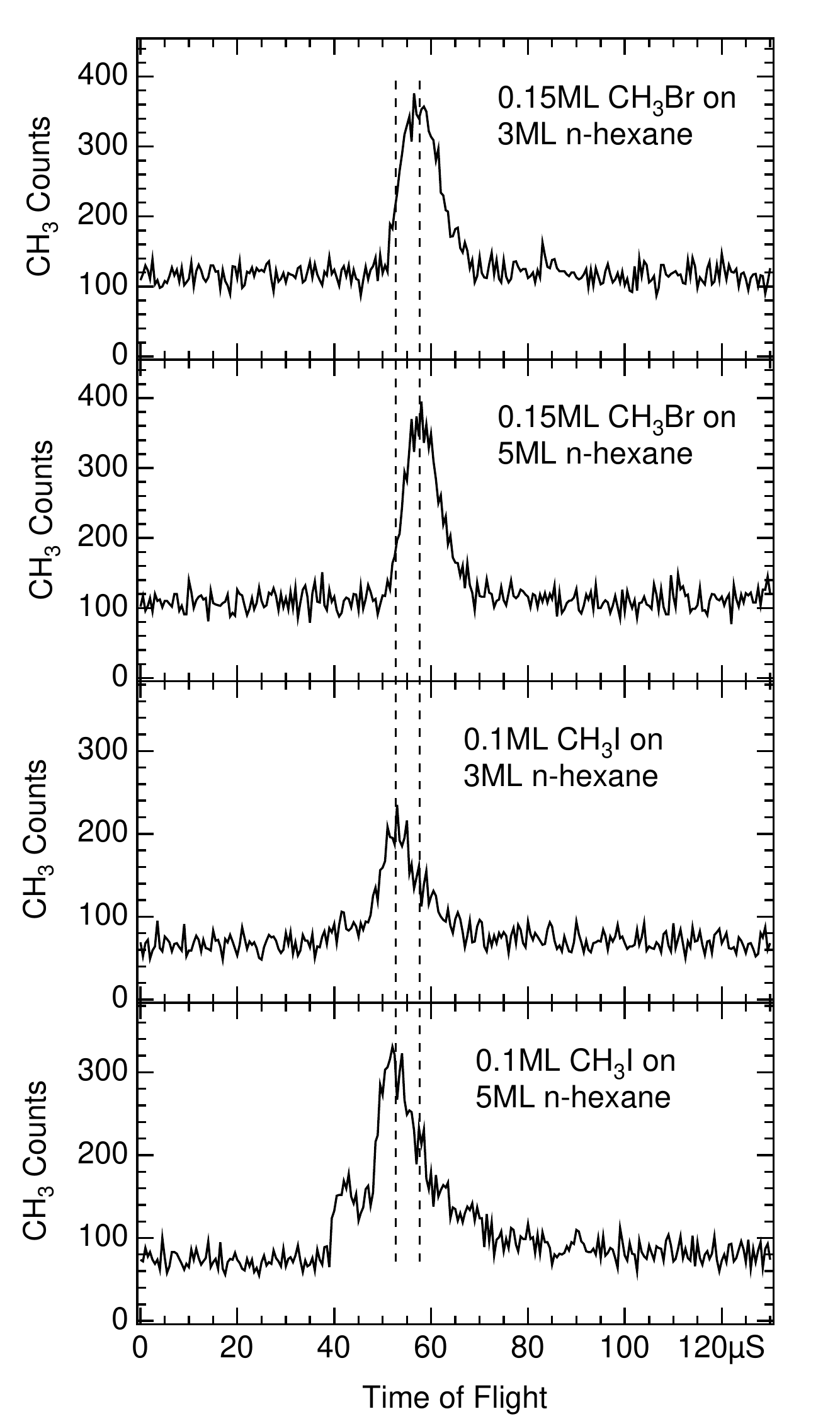}}
\caption{Higher resolution time-of-flight spectra (0.5$\mu$s bins) for submonolayers of CH$_3$Br and CH$_3$I on 3ML and 5ML {\it n}-hexane thin films obtained by summing data three independent scans. These spectra were obtained using unpolarized 248nm light and a Cu(100) substrate. There is no detectable shift in the TOF spectrum for the DEA features for differing {\it n}-hexane film thicknesses, but there is a shift of 5.0$\mu$s between the DEA features for CH$_3$Br and CH$_3$I. }
\label{HiRes_TOF_Comparison}
\end{figure}

The energetics of the dissociation by low energy electrons and the contrast between the case for CH$_3$Br and CH$_3$I are further illustrated by the TOF data shown in Fig. {\ref{HiRes_TOF_Comparison}}. For rapid dissociation via DEA for a free CH$_3$X molecule, the translational energy of the CH$_3$ fragment can be understood from:
\begin{equation*}
\begin{split}
T_{{CH_3}}=\frac{m_X}{m_{CH_3X}}&\left \{   E_{e^-}+EA(X) - D_0(C-X) \right. 
\\ & 
\left. + \Delta E_{solvation}(X^-) - E_{int}(CH_3) \right \}
 \end{split}
\end{equation*}
where the mass ratio accounts for momentum conservation for a dissociating free molecule, $E_{e^-}$ is the kinetic energy of the incident electron, $EA(X)$ is the electron affinity of the halogen atom ($EA(Br)=3.36$eV, $EA(I)=3.06$eV)\cite{CRC_Handbook:2006}, $D_0(C-X)$ is the methyl--halogen bond dissociation energy from the ground state ($D_0(C-Br)=3.05$eV, $D_0(C-I)=2.48$eV)\cite{CRC_Handbook:2006}, $\Delta E_{solvation}(X^-)$ is the additional stabilization energy for the halogen anion at the surface and $E_{int}(CH_3)$ accounts for internal excitation of the CH$_3$ fragment upon dissociation. If we assume that $E_{int}(CH_3)$ is similar for both CH$_3$Br and CH$_3$I and is near zero for the fastest CH$_3$ fragments, then $E_{e^-}+\Delta E_{solvation}\approx $1.0eV gives $T_{CH_3}$ consistent with the observed values. This seems reasonable given that one would expect $\Delta E_{solvation}$ to be of the order of 0.8eV\cite{Michaud:1990}, by analogy with electron scattering measurements for N$_2$ on thin rare gas films and $E_{e^-}\approx$0.2eV is reasonable. Based on these values, the expected TOF {\it differences} for CH$_3$ from CH$_3$Br and CH$_3$I would be roughly 6.0$\mu s$, which is slightly larger than the 5.0$\mu s$ shift we observe in Fig. {\ref{HiRes_TOF_Comparison}}. It would be expected that the solvation energies for the Br$^-$ will be somewhat larger than that for I$^-$, though the precise magnitude of this difference is not known for the halogen anions at a hexane surface-- a difference of $\sim$0.04eV would make the values consistent with the observed flight time differences.

Another interesting observation from Fig. {\ref{HiRes_TOF_Comparison}} is that there is no detectable shift between the TOF spectra from 3ML and 5ML {\it n}-hexane coverages for either CH$_3$Br of CH$_3$I. This implies that $E_{e^-}+\Delta E_{solvation}$ is essentially constant over the range of {\it n}-hexane coverages used for CH$_3$X dissociation. For example, between 3ML and 5ML {\it n}-hexane films the {\it n\/}=1 image state energy rises (from calculations such as that shown in Fig. {\ref{Image_States}}) from -0.40eV to -0.33eV but $\Delta E_{solvation}$ would be expected to decrease\cite{Michaud:1990} by $\sim$0.090eV due to the increased anion to metal surface distance, for an expected change in TOF of +0.5$\mu$s. Within the resolution of our TOF spectra we do not detect any shifts in either the leading edge or the centroid of the TOF distributions. It is also worth noting that while the contribution of higher energy $n$=2 image state electrons would yield significantly faster CH$_3$ fragments (by $\sim 6.8\mu$s), the relative number of these is much smaller\cite{Lingle:1996} than for the $n$=1 image state, which seems to be primarily due to the much smaller spatial overlap of the $n$=2 wavefunction with the population of hot photoelectrons that are transported to the metal interface.

\section{Conclusions}
The photodissociation of CH$_3$Br and CH$_3$I adsorbed on top of thin {\it n}-hexane layers has been found to be caused by hot photoelectrons generated in the metal substrate. The nascent photoelectron energy distribution is modified by the {\it n}-hexane layers and some of these photoelectrons are temporarily trapped at the {\it n}-hexane--vacuum interface in the $n$=1 image state. These image state electrons can interact with the co-adsorbed CH$_3$X and cause dissociation via the DEA mechanism. The CH$_3$ photofragments detected in the TOF experiment have a translational energy distribution that is insensitive to the incident photon energy and the yields of CH$_3$ are indicative of the image state mediated mechanism. This mechanism is likely to be viable for a number of molecular systems, as a wide variety of molecules have the requisite negative electron affinity and bandgap to create such image states, and there are also many molecules which undergo various chemical processes due to interactions with very low energy electrons.

\section*{Acknowledgements}
The author would like to thank the Natural Sciences and Engineering Research Council (NSERC) of Canada for financial support of this work. Thanks also to Jeff Hnybiba, Robert Vogt and Masresha Berhanu, who provided assistance with the experiments as a part of their NSERC Undergraduate Student Research Award tenure in the laboratory. 
\newpage

\end{document}